\documentclass[acmlarge,screen,nonacm=true,11pt]{acmart}

\usepackage{amsmath}
\usepackage{cleveref}
\usepackage{enumerate}
\newtheorem{observation}{Observation}
\newtheorem{theorem}{Theorem}
\newtheorem{lemma}{Lemma}
\usepackage{subcaption}

\title{Parametric Graph Templates: Properties and Algorithms}

\author{Tal Ben-Nun}
\affiliation{%
 \department{Department of Computer Science}
  \institution{ETH Zurich}
}
\email{talbn@inf.ethz.ch}

\author{Lukas Gianinazzi}
\affiliation{%
 \department{Department of Computer Science}
  \institution{ETH Zurich}
}
\email{lukas.gianinazzi@inf.ethz.ch}

\author{Torsten Hoefler}
\affiliation{%
 \department{Department of Computer Science}
  \institution{ETH Zurich}
}
\email{htor@inf.ethz.ch}

\author{Yishai Oltchik}
\affiliation{
\institution{
Next Silicon
}
}
\email{yishai.oltchik@nextsilicon.com}

\begin{document}

\begin{abstract}
	Hierarchical structure and repetition are prevalent in graphs originating from nature or engineering. 
	These patterns can be represented by a class of parametric-structure graphs, which are defined by templates that generate structure by way of repeated instantiation. 
	We propose a class of parametric graph templates that can succinctly represent a wide variety of graphs. 
	Using parametric graph templates, we develop structurally-parametric algorithm variants of maximum flow, minimum cut, and tree subgraph isomorphism. Our algorithms are polynomial time for maximum flow and minimum cut and are fixed-parameter tractable for tree subgraph isomorphism when parameterized by the size of the tree subgraph. By reasoning about the structure of the repeating subgraphs, we avoid explicit construction of the instantiation. 
	Furthermore, we show how parametric graph templates can be recovered from an instantiated graph in quasi-polynomial time when certain parameters of the graph are bounded. 
 Parametric graph templates and the presented algorithmic techniques thus create opportunities for reasoning about the generating structure of a graph, rather than an instance of it.
 \end{abstract}

\maketitle

\clearpage
\setcounter{page}{1}

\section{Introduction}

\begin{figure*}[b]
\vspace{-1em}
\begin{subfigure}[t]{0.31\linewidth}
\centering
	\includegraphics[width=0.9\textwidth]{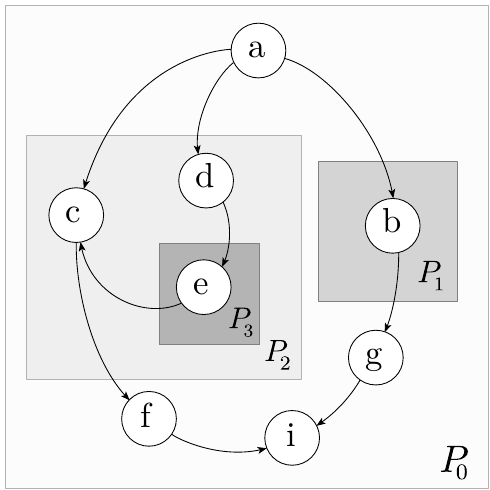}
	\caption{Parametric Graph Template $\mathcal{G}$.} \label{fig:intro-example:template}
\end{subfigure}
\quad
\begin{subfigure}[t]{0.22\linewidth}
\centering
\raisebox{2em}{
	\includegraphics[width=0.55\textwidth]{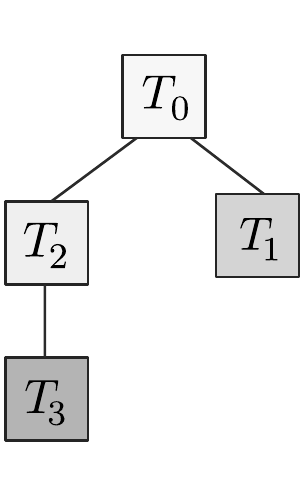}}
	\caption{Template tree of $\mathcal{G}$.} \label{fig:intro-example:tree}
\end{subfigure}
\quad
\begin{subfigure}[t]{0.4\linewidth}
	\includegraphics[width=\textwidth]{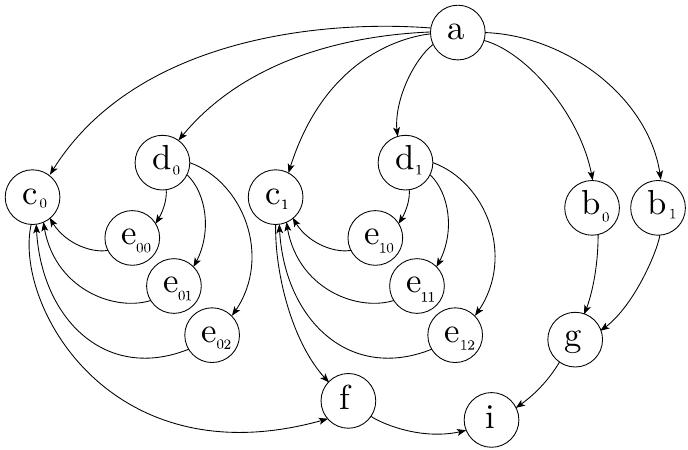}
	\caption{Instantiation of $\mathcal{G}$.} \label{fig:intro-example:instance}
\end{subfigure}
\caption{Illustration of the parametric graph template $\mathcal{G}$ with templates $T_0=\{a, b, c, d, e, f, g, i\}$, $T_1=\{b\},T_2=\{c, d, e\}, T_3=\{e\}$; parameters $P_0=1, P_1=2, P_2=2, P_3=3$; and $h=2$. 
}\vspace{-0.5em}
\label{fig:intro-example}
\end{figure*}

We consider a class of graphs that contain parametric structure. That is, graphs that can be instantiated (i.e., generated) with different numbers of vertices and edges, depending on a set of parameters. As opposed to graphs with parametric edge weights~\cite{DBLP:journals/dam/AnejaCN03, DBLP:journals/mp/GranotMQT12,DBLP:journals/mp/AissiMMQ15,DBLP:conf/stoc/Karger16, DBLP:conf/soda/Erickson10, DBLP:journals/dam/KarpO81}, structural parameterization can affect which subgraphs occur and the number of vertex-disjoint paths between two vertices. 

We focus on a subclass of such parametric graphs, which we call \textit{parametric graph templates}.
A parametric graph template with $k$ parameters $\mathcal{G}=(G, \mathcal{T}, \mathcal{P})$ contains a directed \emph{template graph} $G=(V, E)$ with $n$ vertices $V$ and $m$ edges $E$, a list of \emph{templates} $\mathcal{T}=T_0, T_2, \dotsc, T_{k-1}$, each with $\emptyset \ne T_i \subseteq V$, and a list of positive integer \emph{parameters} $\mathcal{P}=P_0, \dotsc, P_{k-1}$ (see \Cref{fig:intro-example:template}). We can also consider weighted or undirected parametric graph templates by adapting the template graph accordingly.
The templates follow a nested, hierarchical structure, meaning that for every pair of templates they are either disjoint or one of them is strictly contained in the other one (for all $i\neq j$: $T_i \cap T_j = \emptyset$ or  $T_i \subset T_j$ or $T_j \subset T_i$). 
We assume that there is a \emph{root template} $T_0=V$. Hence, the subset relation on the templates induces a \emph{template tree} (see \Cref{fig:intro-example:tree}). We denote its height by $h$. If a template $T$ is contained in another template $T'$ (i.e., $T\subset T'$), then $T$ is a descendant of $T'$ (and $T'$ is an ancestor of $T$). A template $T$ is a parent of $T'$ (and $T'$ is a child of $T$) if $T'$ is the direct descendant of $T$. 
To create an \emph{instantiation} of a parametric graph template $\mathcal{G}$, repeatedly rewrite it as follows (see \Cref{fig:intro-example:instance} for an example). As long as there is more than one template, pick a leaf template $T_i$. For each vertex $v$ in $T_i$ create $P_i$ copies $v_1, \dotsc, v_{P_i}$ called instances of $v$, replacing $v$ in $V$. The set of vertices with the same subscript are called an \emph{instance} of $T_i$. For each edge $e=(u, v)$ with both endpoints in $T_i$, create $P_i$ instances $e_1=(u_1, v_1), \dotsc, e_{P_i}=(u_{P_i}, v_{P_i})$, replacing $e$ in $E$. For each edge $e=(u, v)$ with one endpoint $u$ in $T_i$, create $P_i$ instances $e_1=(u_1, v), \dotsc, e_{P_i}=(u_{P_i}, v)$, replacing $e$ in $E$. Proceed symmetrically for each edge $e=(u, v)$ with one endpoint $v$ in $T_i$. Then, remove the template $T_i$ and its parameter $P_i$. 

Observe that for certain graph problems such as shortest paths, the solution of the problem on the template graph directly corresponds to a solution in the instantiation. For other problems, such as connected components, the solution on the template graph can be easily translated into a solution of the instantiation by scaling it. However, we will see that there are also nontrivial problems, such as maximum $s$-$t$ flows, minimum cuts, and subgraph isomorphism, where a careful study of the structure of the problem is required. For example, even if a tree does not occur as a subgraph of the template graph, it might occur as a subgraph of the instantiation.

\subsection{Motivation}

Graphs that exhibit hierarchical and repeating structures are used in practice in many sub-fields of computer science. Self similarity and repeating structures are fundamental properties of nature and man-made artifacts. Consider the following examples:

\paragraph{Computational Sciences}
In Molecular Biology, macro-molecular structures (such as polymers, proteins) can be represented as graphs of their atomic structure. Hierarchy and repetition are abound in biological structures, especially in self-assembled systems that contain millions of atoms~\cite{Ginsburg2016}, and algorithms such as subgraph isomorphism can be used to find certain structures of interest within measurements.

\paragraph{Programming Languages and Compilers}
Graph analysis is commonplace in the context of compiler analysis~\cite{llvm}. As code is composed of hierarchies of scopes (translation units, functions, statements), hierarchical graph representations~\cite{DBLP:journals/jcss/DrewesHP02} concisely represent programming patterns. Parametric parallelism in applications also appears in compilers~\cite{hpvm} and data-centric programming~\cite{sdfg}. In the latter case, vertices model computation and edges model data movement, and minimum cuts are used to optimize the distribution of such program graphs to multiple compute devices. Lastly, computational DAGs and pebble games~\cite{rbpg} are used to prove I/O lower bounds of algorithms, and contain repeating patterns in computations with loops.

\paragraph{Machine Learning}
Deep Neural Networks (DNNs)~\cite{ddl} are prevalent machine learning models. They usually consist of DAGs containing repeating modules, sometimes called layers or operators, with several levels of hierarchy. The lowermost level (i.e., a single operation such as convolution) contains neurons as nodes that are connected to the neurons of the next layer via weighted edges. While it is infeasible to execute graph algorithms on state-of-the-art models, which approach and exceed trillion parameters~\cite{brown2020language,rajbhandari2020zero}, reasoning about information bottlenecks and expressive power becomes possible in the context of a parametric structure graph.

\paragraph{Networking and Communication}
Hierarchical graphs can be used to succinctly describe a wide range of network topologies used for leading supercomputers.
Hypercube networks~\cite{hypercube}, for example, can be expressed in this form. 
More recent low-diameter hierarchical network topologies~\cite{dragonfly,slimfly,xpander} 
can be described as repeating groups of Hamming grids, cliques, or MMS graphs~\cite{mms98}.
Distributed applications that run on such networks can also be represented by their communication graphs, i.e., which compute nodes communicate and how much.
Determining maximum $s$-$t$ flows and minimal cuts is thus important to the analysis of both network topology, and specific applications and their distribution.

\paragraph{Graph Theory} 
Several classes of general graphs, such as complete $k$-ary trees, series-parallel graphs, or a star of cliques, can be represented by parametric graph templates.

\quad

As the examples demonstrate, the instantiated graphs capture a final state or observation, but do not contain the underlying phenomena that generate it. It is therefore inefficient to reason about the fully instantiated structure instead of arguing about the underlying ``generating'' structure. 
By leveraging the structural patterns, we show that graph theory can be applied directly to the template graph, without instantiating the repeating subunits. 
This translates to algorithm runtimes 
that depend only on the template graph size.

 \subsection{Related Work} 

\paragraph{Graph Grammar}
Graph Grammars~\cite{DBLP:conf/stacs/Courcelle88, DBLP:conf/focs/EhrigPS73, DBLP:journals/jacm/Pavlidis72, DBLP:journals/mst/BauderonC87, DBLP:conf/fct/Engelfriet89} describe a (possibly infinite) language of graphs compactly with a set of construction rules. There is a wide variety of such ways of constructing a graph, differing in expressive power.  A classic problem for graph grammars is to decide whether a graph can be constructed from a given grammar (parsing). In contrast to graph grammars, we are not primarily concerned with expressing an infinite set of graphs, but instead with a succinct representation of a graph and algorithmic aspects of solving graph problems efficiently on this succinct representation.

\paragraph{Hierarchical Graphs} Hierarchical Graphs~\cite{DBLP:journals/jcss/DrewesHP02} model graphs where edges expand to other, possibly hierarchical graphs. They are a variant of context-free hyperedge replacement grammars that incorporate a notion of hierarchy. Graph transformations (i.e., replacing subgraphs within other subgraphs) is explored and tractable~\cite{DBLP:journals/jcss/DrewesHP02}. However, their method does not include parametric replication (i.e., the replacement is context free). Hierarchical graphs also support the notion of replacing a (hyper)edge with a ``variable'' graph. Whilst variables potentially introduce replication/repetition, there are certain restrictions (e.g., a variable can only appear once) that disallow this.
Similarly, \emph{nested graphs}~\cite{DBLP:journals/tois/PoulovassilisL94} allow ``hypernodes'' to represent other nested graphs. The authors focus on the case where a node represents a fixed nested graph. 
Representing parametric graph templates as general hypergraphs or nested graphs, where the hierarchy units are separate, does not suffice for drawing conclusions on the properties of the instantiation based on the non-instantiated structure. 

\paragraph{Edge-Weight Parametric Problems}
Several graph problems have been generalized to the edge-weight parametric case, where edge weights are functions of one or several parameters $\mu_i$. This includes maximum $s$-$t$ flow / minimum $s$-$t$ cut~\cite{DBLP:journals/dam/AnejaCN03, DBLP:journals/mp/GranotMQT12}, (global) minimum cut~\cite{DBLP:journals/mp/AissiMMQ15,DBLP:conf/stoc/Karger16} and shortest paths~\cite{DBLP:conf/soda/Erickson10, DBLP:journals/dam/KarpO81}. The solution is then a piecewise characterization of the solution space. Usually, only linear (or otherwise heavily restricted) dependency of the edge weights on the parameters have been solved. Instead, in our work, we consider a \emph{structurally parametric} generalization of graph problems. This allows us to consider structural problems (such as subgraph isomorphism) in addition to flow and cut problems.

When each edge weight is a linear combination of $k$ parameters $\mu_i$ (i.e., the weight of each edge $e$ is of the form $\sum_{i=1}^{k} c_i(e) \mu_i$ for constants $c_i(e)$), \emph{edge-weight parametric minimum cuts} can be solved in $m^{O(k^2)}$ time~\cite{DBLP:journals/mp/AissiMMQ15}. If the constants $c_i(e)$ and parameters $\mu_i$ are all positive, the runtime is $O(mn^{1+k})$~\cite{DBLP:conf/stoc/Karger16}. Another tractable case is when there is only one parameter $\mu$ and the weight of an edge $e$ takes the form $\sum_{i=1}^{k} c_i(e) \mu^i$. This case takes $O(kn^3 \sqrt{m})$ time to solve~\cite{DBLP:conf/stoc/Karger16}.

For \emph{edge-weight parametric maximum $s$-$t$ cuts}, the problem can be solved in polynomial time when each edge $e$ has weight $\min(c(e), \mu)$ for constants $c(e)$ and a single parameter $\mu$~\cite{DBLP:journals/dam/AnejaCN03}. Granot, McCormick, Queyranne, and Tardella explore other tractability conditions~\cite{DBLP:journals/mp/GranotMQT12}.

In our work, we show how to transform structurally parametric minimum cuts to instances of the edge-weight parametric case. Furthermore, we give conditions to characterize all cuts that can become minimum for any parameters of a parametric graph template. 

\subsection{Problem Statement}

We approach parametric graph templates from an algorithmic perspective. The goal is to solve classical graph problems for \emph{fixed parameters}, but in time that is strongly polynomial in the size of the parametric graph template. We focus on three classes of graph problems: maximum $s$-$t$ flows, minimum cuts, and tractable variants of subgraph isomorphism. These problems have applications in operations research~\cite{10.5555/137406}, network reliability~\cite{DBLP:conf/stoc/Karger95}, graph clustering~\cite{DBLP:journals/ipl/HartuvS00}, electronic circuit design~\cite{DBLP:conf/dac/OhlrichEGS93}, graph mining~\cite{DBLP:conf/icdm/KuramochiK01}, and bioinformatics~\cite{DBLP:journals/jcisd/ArtymiukBGPPRTWW92}.
Moreover, we consider the problem of finding a parametric graph template that instantiates a given graph.

\paragraph{Template Maximum $s$-$t$ Flow}
\emph{An $s$-$t$ flow} $f$ assigns every edge $e$ a nonnegative real flow $f(e) \leq w(e)$. The sum $\sum_{e=(u, v)}f(e) - \sum_{e=(v, w)}f(e)$ is the \emph{net flow} of the vertex $v$. A flow has to have net flow $0$ for all vertices except $s$ and $t$. The value of the flow is the net flow of the source (which equal minus the net flow of the sink). A maximum flow is a flow of maximum value.

The maximum $s$-$t$ flow problem has a natural generalization to parametric graph templates when $s$ and $t$ are vertices in the root template: Instantiate the graph and compute a maximum flow between the only instance of $s$ and the only instance of $t$. 
There are multiple possibilities for how to interpret the case when $s$ and $t$ have multiple instances. One interpretation is as a multiple-source and multiple-target flow problem, where all instances of $s$ are treated as sources and all instances of $t$ as sinks. We call this a \emph{maximum all-$s$-$t$ flow}. 
Another interpretation considers the maximum flow between a fixed instance of $s$ and a fixed instance of $t$. We call this a \emph{maximum single-$s$-$t$ flow}. 

\paragraph{Template Minimum Cut} A cut is a partition of the graph into two disjoint (nonempty) subsets called \emph{sides} of the cut. An edge \emph{crosses} the cut if it has an endpoint in both partitions. The \emph{value} of a cut is the total weight of the crossing edges. A \emph{minimum cut} is a cut of smallest value. We again consider a minimum of a parametric graph template as a cut of its instantiation.

\paragraph{Template Subgraph Isomorphism} A \emph{pattern graph} $H$ is \emph{isomorphic} to a subgraph of a graph $G$ if there is an injective mapping $\phi$ from the vertices of the graph $H$ to the vertices of the graph $G$, where for every edge $(u, v)$ in the graph $H$ there is an edge $(\phi(u), \phi(v))$ in the graph $G$. If a graph $H$ is isomorphic to a subgraph $H'$ of $G$, we say the \emph{pattern} $H$ \emph{occurs in} the graph $G$. $H'$ is an \emph{occurrence} of $H$ in $G$.
A pattern $A$ \emph{occurs in a parametric graph template $\mathcal{G}$} if the graph $A$ occurs in the instantiation of $\mathcal{G}$. 

\paragraph{Template Discovery}
Given an undirected target graph $G'$, the goal in \emph{Template Discovery} is to find a parametric graph template $(G, \mathcal{T}, \mathcal{P})$ whose instantiation is isomorphic to $G'$. We assume that all parameters (except the root's parameter) are at least $2$. Templates with parameter $1$ are not replicated, so they cannot contain meaningful information about repeating substructures.

\pagebreak
\subsection{Results}
We demonstrate that many classical graph problems can be solved asymptotically faster than instantiating the parametric graph template. For many problems, such as maximum $s$-$t$ flow, minimum cut, and tree subgraph isomorphism, it is possible to obtain a runtime that is the similar to the runtime on the template graph. 

For maximum all-$s$-$t$ flow, our algorithms match the runtime of a maximum $s$-$t$ algorithm such as Orlin's $O(mn)$ time algorithm~\cite{DBLP:conf/stoc/Orlin13}. We solve this problem using a technique called \emph{Edge Reweighting}. It observes that scaling the edge weights in the graph template solves the problem.
\begin{theorem}\label{thm:allstflow}
	Computing a maximum all-$s$-$t$ flow in parametric graph template takes $O(mn)$ time.
\end{theorem}

For maximum single-$s$-$t$ flow and minimum cuts, there is an overhead proportional to the height $h$ of the template tree. In addition to Edge Reweighting, we use a technique called \emph{Partial Instantiation}. We observe that a carefully chosen part of the instantiated graph can give sufficient information to extrapolate the result to the rest of the graph. How this part is chosen depends on the problem.
\begin{theorem}\label{thm:singlstflow}
	Computing a maximum single-$s$-$t$ flow in parametric graph template takes $O(mn h)$ time.
\end{theorem}

We combine the insights from the maximum flow problems with observations specific to the structure of minimum cuts of parametric graph templates to match the runtime of classic graph algorithms such as Gawrychowski-Mozes-Weimann's randomized $O(m \log ^ 2 n )$ time algorithm~\cite{DBLP:conf/icalp/GawrychowskiMW20} up to a factor of $h$. In particular, we need to characterize the case where the minimum cut does not correspond to a maximum all-$s$-$t$ cut, without explicitly considering all maximum single-$s$-$t$ cuts.
\begin{theorem}\label{thm:mincuts}
	Computing the minimum cut of an undirected parametric graph template takes $O(m h \log ^ 2 n )$ time (correct with high probability) or $O(mn h + n^2 h \log n)$ time (deterministically).
\end{theorem}

For subgraph isomorphism we provide an algorithm that is linear time for $O(1)$-sized tree patterns. The result uses a technique called \emph{Hierarchical Color Coding}. This technique is a generalization of randomized color coding~\cite{DBLP:journals/jacm/AlonYZ95} to our hierarchical setting. It allows us to solve subgraph isomorphism problems by ``guessing'' how the small subgraph is allowed to behave with respect to the template tree's structure.
\begin{theorem}\label{thm:isomorphs}
Deciding if a rooted tree $A$ of $k$ vertices occurs in a parametric graph template $\mathcal{G}$ rooted at a vertex $v$ (for all vertices $v$) takes $k^{O(k)}m\log n$ time. If all parameters $P_i\geq k$, it takes $2^{O(k)}m\log n$ time.
\end{theorem}

Finally, we show that enumerating all (nontrivial) parametric graph templates that instantiate a graph takes quasi-polynomial time when the number of vertices that are on the boundary of the children templates is bounded by a constant $\beta$. When the graph additionally has bounded treewidth, the runtime is polynomial.

\begin{theorem}\label{thm:autotemplate}
	On a target graph $G'$ with $N$ vertices and $\beta$ is the maximum number $\beta$ of nodes connecting a template to a child template, template discovery takes $N^{  \beta\ \text{\emph{polylog}}( N )}$ time. If $G'$ has treewidth $\tau$, template discovery takes $2^{O(
	\tau^5 \log \tau)}\ N^{ O(\beta\ \log N)}$ time. 
\end{theorem}

\pagebreak
\section{Preliminaries} \label{sec:definitions}

We proceed to introduce definitions, notation, and assumptions that we use throughout this work. 

\paragraph{Template a vertex belongs to.} If a vertex $v$ is in a template $T_i$ and $v$ is in no other template that is a descendant of $T_i$, then $v$ \emph{belongs to} $T_i$. We denote the unique template that $v$ belongs to by $T(v)$. 

\paragraph{Template an edge belongs to}
If both endpoints of an edge belong to a template $T_i$, then this edge belongs to template $T_i$. We denote the number of vertices and edges that belong to a template $T_i$ by $n_i$ and $m_i$, respectively.

\paragraph{Cross-template edges} An edge $(u, v)$ where $u$ and $v$ belong to different templates is \emph{cross-template}.

\paragraph{No Skipping.} We forbid edges that `skip' layers in the template hierarchy. Specifically, if $(u,v)$ is a cross-template edge, then $T(u)$ is a parent or child of $T(v)$. This rule ensures that a path in the graph corresponds to a walk in the template tree.

\paragraph{Boundary Vertices}

Consider a vertex $u$ and $v$ where $T(v)$ is a parent of $T(u)$. If there is an edge from $u$ to $v$ or from $v$ to $u$ in the template graph, then $v$ is a \emph{boundary vertex} of $T(u)$.

\paragraph{Template graph of a template.}
The subgraph of the template graph $G$ induced by a template $T_i$ is called the \emph{template graph of} $T_i$. 

\paragraph{Instance tree.} We extend the nomenclature of templates to instances. The template hierarchy can be transferred onto the instances, where an instance $I$ is a descendant of an instance $I'$ if the template $T$ that instantiated $I$ is a descendant of the template $T'$ that instantiated $I'$. Similarly, we extend the notions of ancestor, parent, and child to the instances, creating an \emph{instance tree}. Two instances that have the same parent instance are \emph{siblings}.

If a vertex $v$ is contained in an instance $I$, but it is not contained in any other descendant of $I$, the vertex $v$ belongs to the instance $I$. 
If $b_i$ is an instance of a boundary vertex $b$ of a template $T$, then $b_i$ is a boundary vertex of the instance that $b_i$ belongs to. The instance of the root template is the \emph{root instance}. For a vertex $v$ in the instantiation, we write $T(v)$ for the template of the instance that $v$ belongs to.

\paragraph{Isomorphism.} Two parametric graph templates $\mathcal{G}_1$ and $\mathcal{G}_2$ are \emph{isomorphic} if they instantiate isomorphic graphs. Note that $\mathcal{G}_1$ and $\mathcal{G}_2$ can have different parameters, templates,  and their template graphs need not be isomorphic for them to instantiate isomorphic graphs.

\paragraph{Cycles}

Acyclic graphs are easier to handle for many algorithmic problems. In parametric graph templates, we consider two different notions of what constitutes a cycle.
The simplest notion of cycles comes from considering cycles in the template graph. If it does not contain any cycles, then the instantiation does neither (and vice versa). 

\paragraph{Template-acyclic}
A path $p_1, \dotsc p_k$ in the template graph that contains three vertices $p_i, p_j, p_k$ with $i<j<k$ and $T(i)=T(k)$ but $T(i) \neq T(j)$ is a template-cycle. We say a parametric graph template is \emph{template-acyclic} if it does not contain a template-cycle. This notion is incomparable to the notion of acyclic parametric graph templates. 
There are acyclic parametric graph templates that are not template-acyclic and a template-acyclic graph can have cycles (as long as all the vertices in the cycle belong to the same template).

\section{Algorithm and Proof Techniques}

In this section, we present the main algorithmic and proof tools that we later apply to solve the aforementioned graph problems. The techniques are surprisingly simple, but their application requires a careful analysis to prove correct, which we defer to \Cref{sec:stflow,sec:mincuts,sec:tree-si}.

\subsection{Edge-Reweighting}\label{sec:tech:edge-reweight}

\begin{figure}[b]\label{fig:edge-reweighting}
\includegraphics[width=0.72\linewidth]{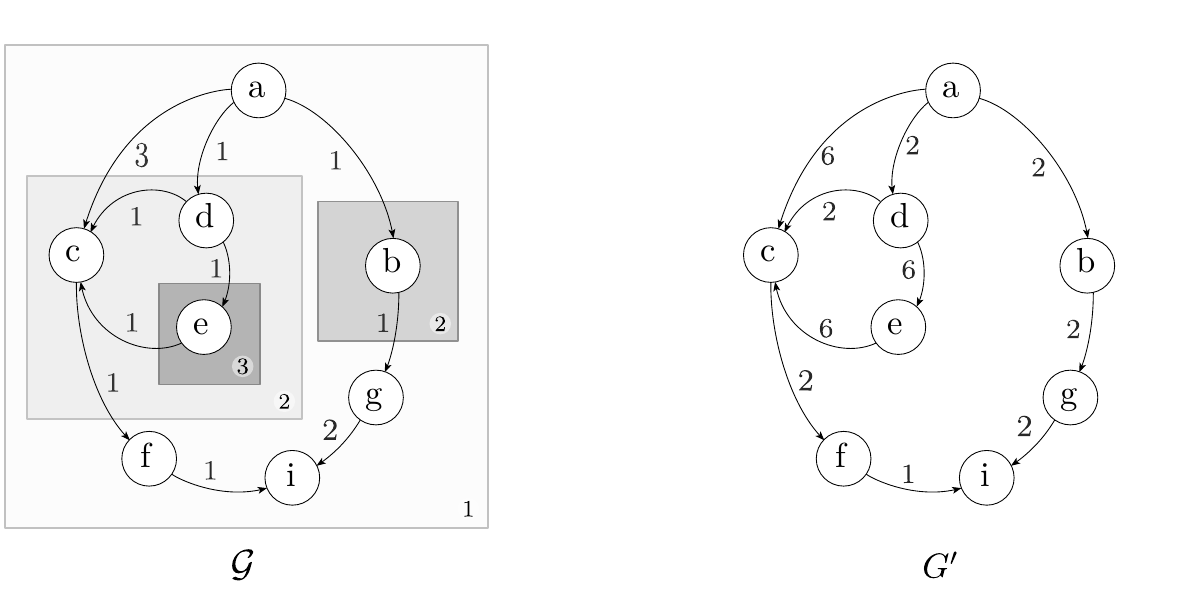}
\caption{Edge Reweighting turns a parametric graph template $\mathcal{G}$ into a graph $G'$ by scaling the weights of each edge in its template graph by all the parameters of the templates that contain an endpoint of the edge. As we will see in \Cref{sec:stflow}, this preserves maximum all $s$-$t$ flow values. }\label{fig:tech:reweight}
\end{figure}

The simplest way to solve a problem on parametric graph templates is to show how it relates to a problem on the template graph with \emph{scaled weights}. We will prove in \Cref{sec:stflow} that for maximum all-$s$-$t$ this technique can be directly applied, and for maximum single $s$-$t$ flow it can be applied to a transformed parametric graph template. Also for minimum cuts, scaling the edge weights solves parts of the problem, as we show in \Cref{sec:mincuts}. Intuitively, Edge Reweighting works whenever all instances of a vertex behave symmetrically. See \Cref{fig:tech:reweight} for an example of Edge Reweighting.

\paragraph{Algorithm}
Transform the parametric graph template $\mathcal{G}=(G,\mathcal{T}, \mathcal{P})$ with edge weights $w$  
into a graph $G'$ with edge weights $w'$. The \emph{reweighted graph} $G'$ has the same vertex and edge set as the template graph $G$, but the weights are scaled as follows:
Multiply the weight of an edge in the template graph by the product of the parameters of the templates that contain at least one endpoint of the edge. That is, let $I(e)$ be the index set of all templates that contain at least one of the endpoints of $e$. Then, the weight of $w'(e)$ is $w(e) \prod_{i\in I(e)} P_i$. To implement this in linear time $O(m)$, precompute in a pre-order traversal of the template tree for each template the product of all the ancestors' parameters. 

We prove the following relation of Edge Reweighting to maximum $s$-$t$ flows in \Cref{sec:stflow}.
\begin{lemma}\label{thm:edge-reweighting}
Edge reweighting of a parametric graph template $\mathcal{G}$ produces a reweighted graph $G'$ where the value of the maximum $s$-$t$ flow of $G'$ equals the value of the maximum all-$s$-$t$ flow of $\mathcal{G}$.
\end{lemma}

\subsection{Partial Instantiation}\label{sec:tech:partial-instantiation}

\begin{figure}[b]
\includegraphics[width=0.55\linewidth]{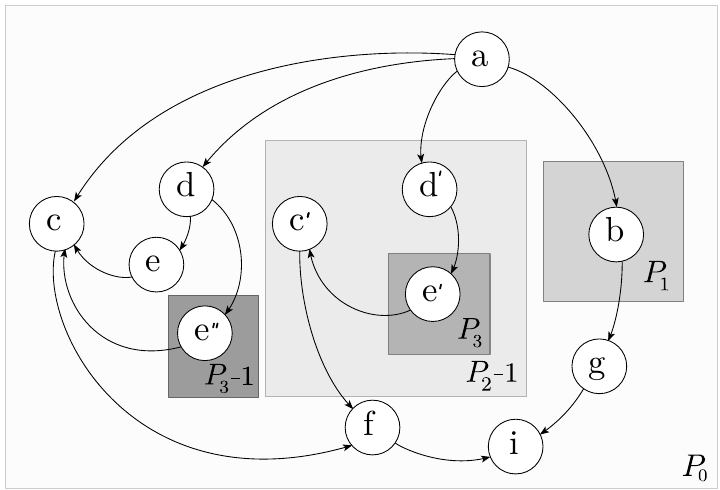}
\caption{After running Upwards Partial Instantiation from $e$ on the parametric graph template $\mathcal{G}$ from \Cref{fig:intro-example:template}, the vertex $e$ is in the root template. The transformed parametric graph template instantiates the same graph.}
\label{fig:tech:part}
\end{figure}

The technique of \emph{partial instantiation} revolves around instantiating only part of the parametric graph template, depending on the problem at hand. The goal is to choose the partial instantiation such that the remaining problem is solvable by using the symmetry of the problem (for example using edge-reweighting). 
The partial instantiation technique can be seen as an example of the more general technique of \emph{retemplating}. The intuition of retemplating is that in certain cases, it suffices to change the representation of the parametric graph template into another isomorphic parametric graph template to significantly simplify the problem at hand.

Next, show how to move a single vertex $s$ from deep in the template tree to the root, without changing the instantiated graph. This is useful for solving maximum $s$-$t$ problems when $s$ (or $t$) belongs to a template that is deep in the template tree (See \Cref{sec:maxflow-acyc}). We call this following technique \emph{Upwards Partial Instantiation} from $s$. For simplicity, let us start with the special case of template-acyclic graphs.

In a template-acyclic parametric graph template, once a path goes from an instance of a template $T_i$ to its parent, it never enters another instance of $T_i$ again. This property implies that, when considering the reachable subgraph from a vertex that is an instance of $s$, we can simply ``merge'' $T(s)$ and all the templates that are ancestors of the template $T(s)$ in the template tree. Formally, this corresponds to deleting $T(s)$ and all the templates that are ancestors of $T(s)$ (except the root) from the parametric graph templates' list of templates.

If the parametric graph template has template-cycles, our goal remains to transform the parametric graph template into an equivalent graph where a particular instance of a vertex $s$ is in the root template. See \Cref{fig:tech:part} for an illustration of Upwards Partial Instantiation.

\paragraph{Algorithm}
Repeat the following until all templates from $T(s)$ to the root have parameter $1$:
\begin{enumerate}
	\item Consider the topmost template $T$ that contains $s$ and has parameter greater than $1$. Let $P_s$ be the number of instances of the template $T$.  
	\item Instantiate the template $T$ twice. Create a new parametric graph template that has the two instances as templates, where the first template has parameter $1$ and the second template has parameter $P_s-1$. The vertices in the second template are relabeled ($s$ is in the one with parameter $1$).
\end{enumerate}
Now, merge $T(s)$ and all the templates that are ancestors of $T(s)$, leaving $s$ in the root template. 

Because this process performs the same rewriting of the parametric graph template as instantiation, just in a different order and stopping early, this process creates an isomorphic parametric graph. Every iteration adds at most $n$ vertices and $m$ edges, there are at most $x$ iterations. We conclude that:
\begin{observation}
	Upwards Partial Instantiation from $s$ produces an isomorphic parametric graph template with at most $x$ additional templates and $O(nx)$ vertices and $O(mx)$ edges in the template graph, where $x$ is the depth of $s$ in the template tree.
\end{observation}

\subsection{Hierarchical Color Coding}\label{sec:tech:color}
\begin{figure}[b]\label{fig:tech:colorcode}
\includegraphics[width=0.65\linewidth]{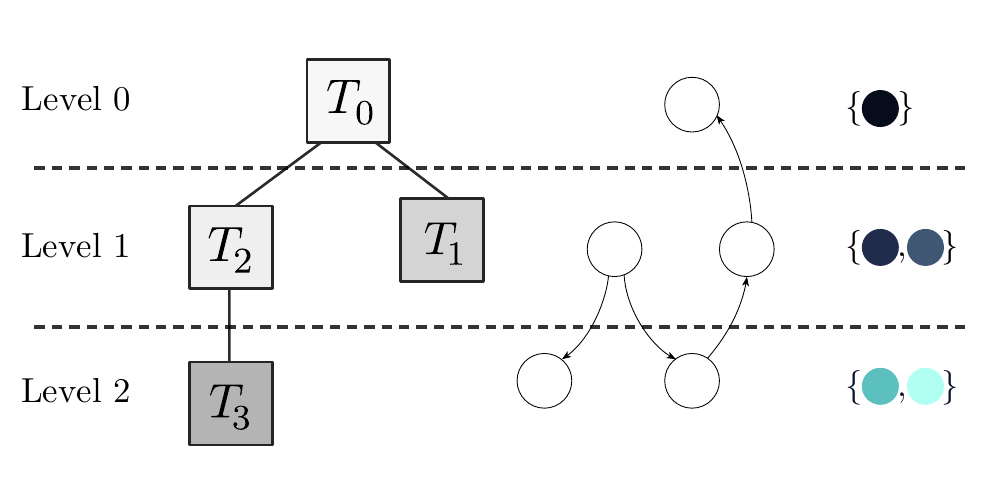}
\caption{Hierarchical Color Coding divides the pattern graph among the levels of the template tree. Then, it assigns a number of colors equal to the number of vertices in the level. Randomly color-coding the template graph with these colors allows us to correctly match patterns more efficiently.}\label{fig:tech:colorcode}
\end{figure}

Color coding~\cite{DBLP:journals/jacm/AlonYZ95} is a randomized technique to find isomorphic subgraphs. The idea is to assign each vertex in the graph a random color (from a set of $k$ possible colors, where $k$ is the number of vertices in the pattern graph) and then restrict our attention to occurrences of the pattern graph where every vertex in the occurrence has a different color. Such occurrences are called \emph{colorful}. This means that it suffices to keep track of the set of colors which have been used by a partial match (instead of the vertices of all partial matches).
Note that when we speak of \emph{colorings}, we do not speak of proper colorings necessarily, that is, neighboring vertices can have the same color. 

We cannot directly apply color-coding on parametric graph templates, because a template can be instantiated multiple times to match a pattern. Hence, some colors must be allowed to repeat. The main insight consists in characterizing the condition when colors of vertices are allowed to overlap. It turns out that we need to be able to distinguish the depth in the template tree at which a color occurs and allow colors to overlap in certain templates that depend on a combination of the structure of the pattern and the template tree. 

An additional challenge is that we do not want to use more colors than the number of vertices of the pattern (in order to keep the state space small). To be able to distinguish which color belongs to which depth of the template tree, we use the same set of colors only in templates that are at the same distance to the root of the template tree. Moreover, we guess the way in which the pattern is mapped onto the template tree. This still allows us to decide if colors come from descendants or ancestors in the template tree when we are combining subgraphs at a certain vertex, whilst keeping the set of possible colorings small. See \Cref{fig:tech:colorcode} for an example of the Hierarchical Color Coding.

\paragraph{Algorithm} Given a parametric graph template $\mathcal{G}$ and pattern graph $A$, proceed as follows:
\begin{enumerate}
	\item Guess how the pattern $A$ maps onto the template tree. This process assigns each vertex in the pattern $A$ a \emph{level number} between $0$ and $h-1$. Next, we describe a way to enumerate all possible choices. Consider a rooted spanning tree of $A$. Choose for the root any arbitrary level number. Then, recursively, choose for each child $v$ of the root $r$ its level number. Say the current root $r$ has level number $l$. There are three choices for the level number of the child $v$:
	\begin{itemize}
		\item $l$ \enspace (meaning that $T(v)=T(r)$),
		\item $l+1$ \enspace (meaning that $T(v)$ is a child of $T(r)$), or
		\item $l-1$  \enspace (meaning that $T(v)$ is a parent of $T(r)$).
	\end{itemize}
	\item For each possible choice from step (1), produce a random coloring as follows. Count how many vertices in the tree $A$ there are for each level number and denote the number of vertices with level number $i$ with $l_i$. Then, partition the set of $k$ colors such that $l_i$ colors are assigned to level $i$. Now, we are ready to color the vertices of the graph: For a vertex that belongs to a template of distance $i$ to the root template, choose a color uniformly at random from the colors that are assigned to level $i$. 
\end{enumerate}

\begin{observation}\label{obs:colorcode1}
There are at most $h3^{k-1}$ possibilities of how the spanning tree of $A$ maps onto the template tree.	
\end{observation}
\begin{proof}
	There are $h$ possible start points in the tree and for every of the $k-1$ edges in the chosen spanning tree there are at most $3$ choices. 
\end{proof}

We extend the coloring conceptually to any instantiation of the parametric graph template by assigning each instance $v'$ of a vertex $v$ the color of $v$. An occurrence of a subgraph $A$ is \emph{semi-colorful} if vertices have distinct colors except for instances of the same vertex.

\begin{observation}\label{obs:colorcode2}
For at least one of the choices of how the pattern maps onto the template tree, the probability that there is a semi-colorful occurrence of $A$ is at least $e^{-k}$.
\end{observation}
\begin{proof}
Consider an occurrence of $A$. Map its vertices back to the vertices in the template graph that instantiated them. This yields a set of at most $k$ vertices. If this set of vertices is assigned distinct colors, then $A$ is semi-colorful. The probability that this occurs is at least $k!/k^k$, which is at least $e^{-k}$~\cite{DBLP:journals/jacm/AlonYZ95}, by the series expansion of the exponential. 
\end{proof}
Hence, $O(e^{k} \log n)$ repetitions of the color-coding suffice for there to be a semi-colorful occurrence with high probability. 

This concludes our overview of the algorithmic techniques. We will now exemplify how to apply them to specific problems in \Cref{sec:stflow}, \Cref{sec:mincuts}, and \Cref{sec:tree-si}. In \Cref{sec:autotemplate}, we will show how to discover all parametric graph templates that instantiate a given graph.

\pagebreak
\section{Template Maximum Flows} \label{sec:stflow}


Next, we turn to the first algorithmic question on parametric graph templates. Our goal here is to solve the maximum $s$-$t$ flows problem on a parametric graph template \emph{without explicitly instantiating it}. Instead, the goal is to get a runtime that is polynomial in the size of the graph template. We will prove the edge reweighting \Cref{thm:edge-reweighting} (which directly gives an algorithm for maximum all-$s$-$t$ flows) and show how partial instantiation can be used to solve maximum single-$s$-$t$ flows.

We will approach the problem by considering (in \Cref{sec:maxflow:root}) the case where $s$ and $t$ are in the root template first. Then, we show how to reduce both the maximum all-$s$-$t$ flow and the maximum single-$s$-$t$ flow problem to an instance of this simpler problem. Throughout, we assume that all vertices are reachable from $s$ and can reach $t$, as otherwise they cannot carry flow. 

In the template-acyclic case, the maximum single-$s$-$t$ flow is trivially zero \emph{except} when $s$ and $t$ are in the same instance of the least common ancestor of $T(s)$ and $T(t)$ in the template tree. Therefore, in the acyclic case it makes sense to restrict our attention to this case where the flow is not trivially zero. In the case where there are template-cycles, it matters which instances of $s$ and $t$ are picked. These can be identified by numbering the instances they belongs to.

The idea is to use Edge Reweighting, because an edge that intersects template $T_i$ can be used $P_i$ times and can therefore be used to carry $P_i$ times the amount of flow. This observation holds as long as $s$ and $t$ are in the root template or if we consider the maximum all-$s$-$t$ flow problem.

Hence, it suffices to run Edge Reweighting and use a maximum $s$-$t$ flow on the scaled graph $G'$, such as Orlin's Algorithm~\cite{DBLP:conf/stoc/Orlin13}. The reweighting can be done in linear time $O(m)$. 

The linear programming \emph{dual} of a maximum $s$-$t$ flow is a \emph{minimum $s$-$t$ cut}~\cite{DBF55}. We will use strong duality~\cite{chvatal1983linear} in our proof of \Cref{thm:edge-reweighting}, which means that it suffices to identify an $s$-$t$ flow and a minimum $s$-$t$ cut of equal value to prove that they are optimal. We argue that Edge 
	Reweighting preserves the value of the dual minimum all-$s$-$t$ cut. Hence, ii also preserves  the maximum all-$s$-$t$ flow value. 

\subsection{Source and Sink Belong to the Root Template}\label{sec:maxflow:root}

We begin with the case where the source $s$ and the sink $t$ are in the root template. In \Cref{sec:allstflow}, we will see that the other case can be easily reduced to this case using instance merging. If $s$ and $t$ are in the root template (which is repeated only once), then a maximum single-$s$-$t$ flow equals a maximum all-$s$-$t$ flow and we call it a maximum $s$-$t$ flow for short.
 The proof uses strong duality and induction on the number of templates.

\begin{lemma}\label{lem:stcut-structure}
In a parametric graph template $\mathcal{G}=(G,\mathcal{T}, \mathcal{P})$, if $s$ and $t$ are in the root template, there is a minimum $s$-$t$ cut of the instantiation of $\mathcal{G}$ where every instance of every vertex is on the same side of the cut.
\end{lemma}

\begin{proof}
The proof is by induction on the number of templates in the parametric graph template. If the parametric graph template has only a single template, then (since $s$ and $t$ must be in this template) the claim is trivial because the root is repeated only once, by assumption.

Otherwise, let $C=(V_s, V_t)$ be a minimum all-$s$-$t$ cut of the parametric graph template $\mathcal{G}$ (i.e., $V_s$ contains the vertices assigned to $s$ and $V_t$ those assigned to $t$).
Consider an arbitrary template $T_i$ that is a \emph{child of the root template} and its graph template $G_i$.
The sets $B_s$ and $B_t$ contain the boundary vertices of $T_i$ that are in $V_s$ and $V_t$, respectively. 

If either of the sets $B_s$ or $B_t$ is empty, then it follows immediately that all instances of the vertices that are in $T_i$ are in the same part of the cut $C$ (namely on the side of $s$ if the set $B_t$ is empty and vice versa). 

Otherwise, merging all vertices in $B_s$ into a vertex $s'$ and merging all vertices in $B_t$ into a vertex $t'$ does not change the value of the minimum $s$-$t$ cut in $\mathcal{G}$. Moreover, if the merged parametric graph template has a minimum $s$-$t$ cut that puts every instance of every vertex on the same side of the cut, then so does the original parametric graph template (because $B_s$ and $B_t$ contain only vertices that belong to the root template and we can ``undo'' the merging). We thus further assume w.l.o.g. that $B_s$ and $B_t$ contain a single vertex named $s'$ and $t'$, respectively. Note that since these vertices belong to the root template, the vertices $s'$ and $t'$ coincide with their only instances.

Every minimum $s$-$t$ cut must separate $s'$ from $t'$ in the subgraph $H$ given by the instances of $T_i$ and the vertices $s'$ and $t'$ (but without a potential edge from $s'$ and $t'$). We use our induction hypothesis to show that there is a minimum $s'$-$t'$ cut in this subgraph that puts all instances of a vertex on the same side of the cut. 

We construct a parametric graph template $\mathcal{G}''$ such that a maximum $s'$-$t'$ flow in $\mathcal{G}''$ can be extended to a flow for the graph $H$. The parametric graph template $\mathcal{G}''$ has the following template graph: take the subgraph of $G$ induced by $T_i$ together with its boundary vertices, then delete any edges going between $s'$ and $t'$. The boundary vertices $s'$ and $t'$ and all vertices that belong to $T_i$ are put into the root template of $\mathcal{G}''$ (which has parameter $1$). Moreover, $\mathcal{G}''$ has the templates and parameters of the descendants of $T_i$ in $G$. The parametric graph template $\mathcal{G}''$ contains at least one template less than $G$. Hence, by induction, there is a minimum $s'$-$t'$ cut $C''$ of $\mathcal{G}''$ that puts all instances of the same vertex into the same partition. 

Let $f''$ be the dual maximum $s'$-$t'$ flow corresponding to $C''$ in $\mathcal{G}''$ of value $\mu$. Now, we construct a $s'$-$t'$ flow $f$ in $H$ and show it is maximum. Along each instance of each edge $e$ that intersects $T_i$ we send $f''(e)$ flow. The capacity constraint on the flow is trivially satisfied. The conservation constraint on the flow is satisfied because in the instantiated graph, the total flow going in and out of an instance of $v$ is the same as for vertex $v$ for $f''$ in $\mathcal{G}''$. The value of the flow $f$ is $P_i \cdot \mu$.

Now, consider the cut $C'$ where we put every instance of a vertex $v$ in $T_i$ on the same side as $v$ is in $C''$. The value of this cut is $P_i \cdot \mu$. By strong duality, this shows that $C'$ is a minimum $s'$-$t'$ cut in the graph $H$. By construction, this cut puts every instance of every vertex on the same side of the cut.

We conclude that all children of the root template can be cut such that every instance of the same vertex is in the same part of the cut. Because the root has a single instance, the statement follows for the root as well.
\end{proof}

\Cref{lem:stcut-structure} shows us how to construct an $s$-$t$ cut $C'$ in the transformed graph $G'$ from an $s$-$t$ cut $C$ in $\mathcal{G}$ of the same value. Together with the other (easier) direction of the proof, this shows that the transformed graph $G'$ has the same maximum $s$-$t$ flow.

\begin{lemma}\label{lem:stcut-reweighting}
	If a parametric graph template $\mathcal{G}$ has a minimum $s$-$t$ cut of value $\mu$ and $s$ and $t$ are in the root template, then the graph $G'$ constructed by edge reweighting has a minimum $s$-$t$ cut of value $\mu$.
\end{lemma}
\begin{proof}

Any cut in the reweighted graph $G'$ corresponds to a cut of the same value in the parametric graph template $\mathcal{G}=(G,\mathcal{T}, \mathcal{P})$: Put every instance of a vertex into the partition that it has in the cut of $G'$. Since every edge $e$ is cut exactly $\prod_{i\in I(e)} P_i$ times, this shows that the value of the minimum $s$-$t$ cut of the graph $G'$ is at least the value of the minimum all-$s$-$t$ cut of the parametric graph template $(G,\mathcal{T}, \mathcal{P})$.

It remains to show that the minimum $s$-$t$ cut of the reweighted graph $G'$ is at most the value of the minimum $s$-$t$ cut of the parametric graph template $\mathcal{G}$.
By \Cref{lem:stcut-structure}, there is a minimum $s$-$t$ cut $C$ of $\mathcal{G}$ that puts every instance of every vertex on the same side of the cut. Now, we construct a cut $C'$ of $G'$ from this cut $C$ by putting every vertex $v$ in $G'$ on the same side of the cut as all the instances of $v$ are in $C$. 
The cut $C'$ has the same value $\mu$ because every instance of an edge $e$ in $\mathcal{G}$ is crossing $\prod_{i\in I(e)} P_i$ times, which is the amount by which we scaled the weight of edge $e$ in $G'$. 
\end{proof}

\subsection{Instance Merging}\label{sec:tech:instance-merging}

Next, we show how to merge all instances of a vertex $v$ in a parametric graph template by transforming it into parametric graph template of almost the same size (the overhead is an additive $O(nh)$). We will use this technique to reduce the general case for maximum all-$s$-$t$ flow to the case where $s$ and $t$ are in the root.

\begin{figure}
\includegraphics[width=0.35\linewidth]{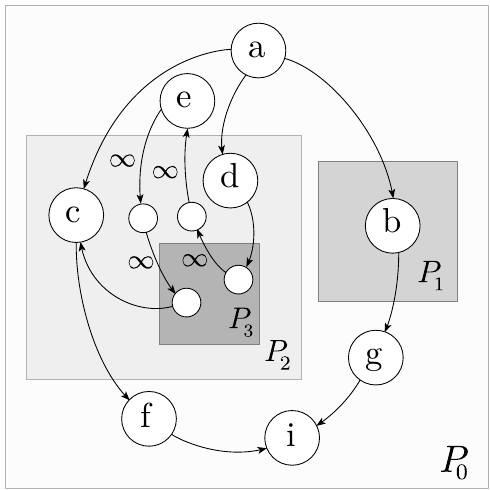}
\caption{After running Instance Merging on the graph $\mathcal{G}$ from \Cref{fig:intro-example:template} with vertex $e$, the vertex $e$ is pushed into the root template. This introduces dummy nodes (drawn smaller and without labels) and dummy edges (of weight $\infty$). The instantiation of the transformed parametric graph template is (after contracting all dummy edges) isomorphic to the graph we get by merging all instances of $e$ in the instantiation of $\mathcal{G}$.}\label{fig:tech:im}
\end{figure}

The idea is that merging all instances of a vertex $s$ is akin to moving the vertex from the template $T(s)$ it belongs to into the root template (so that it belongs to the root template). The \emph{no skipping rule} only allows edges to go from parent templates to children templates (or vice versa), we need to introduce \emph{dummy edges} and \emph{dummy vertices} along the way. The dummy edges have $\infty$ weight. An original edge $(u, s)$ will be transformed into a path $u, d_1, \dotsc, d_k, s$ for dummies $d_1, \dotsc, d_k$ (symmetrically for an edge $(s, u)$). See \Cref{fig:tech:im} for an example.
 
\paragraph{Algorithm} Given a parametric graph template $\mathcal{G}$ and a vertex $s$, repeat the following until $s$ is in the root template:
\begin{enumerate}
	\item For any cross-template edge $(u, s)$, introduce a dummy vertex $d$ in the template $T(s)$ that $s$ belongs to. Replace the edge $(u, s)$ by two edges $e_1=(u, d)$ and $e_2=(d, s)$. The weight of the edge $e_1$ is the same as the weight of the edge $e$, but the weight of the edge $e_2$ is set to $\infty$. Proceed symmetrically for any cross-template edge $(s, u)$.
	\item Move the vertex $s$ from the template $T(s)$ to the parent of the template $T(s)$ (i.e., remove $s$ from the set $T(s)$). 
\end{enumerate}

\begin{observation}\label{lem:tech:instance-merge}
	Instance Merging($\mathcal{G}$, $s$) produces a parametric graph template $\mathcal{G'}$ whose instantiation is, after merging all instances of dummy edges of weight $\infty$, isomorphic to the graph that we get by instantiating the original parametric graph template $\mathcal{G}$ and merging all instances of $s$. Instance Merging($\mathcal{G}$, $s$) adds at most $d(s) \cdot h$ vertices and edges, where $d(s)$ is the degree of the vertex $s$ in the template graph. 
\end{observation}
\begin{proof}
	In the template graph of $\mathcal{G'}$, there is a path consisting of $\infty$-weight edges from $s$ to every neighbor of $s$ in the template graph $G$ of $\mathcal{G}$. There are no other $\infty$ weight edges. Hence, there also is such an $\infty$-weight path in the instantiation of $\mathcal{G'}$ to every instance of every vertex that is a neighbor of $s$ in $G$. Contracting these paths gives a graph where the vertex $s$ has an edge to all instances of neighbors of $s$ in $G$, which is the same graph that we get by instantiating the original parametric graph template $\mathcal{G}$ and merging all instances of $s$. 
\end{proof}

\subsection{Maximum All-$s$-$t$ Flow}\label{sec:allstflow}

To solve maximum all $s$-$t$ Flow, all we would need to do is use Instance Merging on $s$ and then on $t$ to ensure that they are both in the root template. Then, we could use the edge reweighing \Cref{lem:stcut-reweighting}. This approach would cost  $O( n m + n^2 h)$ time. We can avoid this overhead by observing that edge reweighting works directly for maximum all-$s$-$t$ Flow (even when $s$ and $t$ are not in the root template), proving \Cref{thm:edge-reweighting} and \Cref{thm:allstflow}.

\begin{proof}[Proof Of \Cref{thm:edge-reweighting}]
Instance Merge $s$ and then $t$ in $\mathcal{G}$ to produce a parametric graph template $\mathcal{G'}$. By definition, all instances of $s$ (and $t$ respectively) must be on the same side of a minimum all-$s$-$t$ cut, this parametric graph template $\mathcal{G}'$ has the same minimum all-$s$-$t$ cut value as the original parametric graph template $\mathcal{G}$. Edge reweighting $\mathcal{G'}$ gives us a graph $\hat G$. From \Cref{lem:stcut-reweighting} we know that a minimum $s$-$t$ cut of $\hat G$ corresponds to the minimum all-$s$-$t$ cut of $\mathcal{G'}$ (which puts all instances of the same vertex on the same side of the cut).

An $\infty$-weight edge never crosses a minimum $s$-$t$ cut and therefore such dummy edges (introduced by the instance merging) from $\hat G$ can be contracted, yielding a graph $G'$. This graph $G'$ is the same graph that we get from edge reweighting the original parametric graph template $\mathcal{G}$ (using \Cref{lem:tech:instance-merge}).
\end{proof}

\subsection{Maximum Single-$s$-$t$ Flow}\label{sec:maxflow-acyc}

We give a partial instantiation and edge reweighting approach to maximum single-$s$-$t$ flow. For there to be a flow through some instance, it must lie along an $s$-$t$ path. Hence, we can use Upwards Partial Instantiation twice to ensure that $s$ and $t$ lie in the root template. Then, we use Edge Reweighting.

\paragraph{Algorithm}
\begin{enumerate}
	\item Perform upwards partial instantiation from $s$ (see \Cref{sec:tech:partial-instantiation}).
	\item Perform upwards partial instantiation from $t$.
	\item Construct an edge-reweighted graph $G'$ (see \Cref{sec:tech:edge-reweight}).
	\item Run a maximum $s$-$t$ flow algorithm on the partially instantiated and reweighted graph $G'$.
\end{enumerate} 

Note that if the parametric graph template is template-acyclic, then partial instantiation does not increase the size of the template graph. We can conclude:
\begin{corollary}\label{cor:stflow-single}
On a template-acyclic graph, Maximum Single-$s$-$t$ Flow takes $O(mn)$ time.
\end{corollary}

\section{Template Minimum Cuts}\label{sec:mincuts}

We turn our attention to undirected graphs and consider the flow-related problem of minimum cuts. 
%
%
Using $s$-$t$ flow computations to compute a minimum cut would not be very efficient. A minimum cut could be expressed as the largest maximum single-$s$-$t$ flow over all pairs $s$ and $t$, yielding an $O(mn^2h)$ algorithm using \Cref{thm:singlstflow}.

By carefully studying the structure of minimum cuts in the parametric case, we can avoid doing any maximum $s$-$t$ flow computations and get a running time that is only an $O(h)$ factor away from the runtime of minimum cuts on the template graph. We will use our Edge Reweighting \Cref{thm:edge-reweighting}.

In our proof of \Cref{thm:mincuts}, we distinguish between two cases. The first case is when for every instance its boundary vertices are on the same side of a minimum cut. In the first case, the structure will be quite different than for the maximum all-$s$-$t$ flows and does not relate to the Edge Reweighting algorithm. The second case is when there is an instance that has boundary vertices on different sides of a minimum cut. The analysis of the second case is naturally related to the Edge Reweighting algorithm and the structural $s$-$t$ cut \Cref{lem:stcut-structure}. If there is an instance of the parent of a template $T$ where the boundary vertices of $T$ are on different sides of the cut, we say the \emph{template $T$ crosses the cut}.
We prove the two cases separately in \Cref{sec:mincut-notemplatecut} and \Cref{sec:mincut-templatecut}. The algorithm simply computes both cases and returns the minimum of the two.

We assume that every template has a single connected component. Otherwise, we can split the template into its connected components (creating a new template for each component) without changing the result or the height of the template tree.

\subsection{No Template Crosses the Minimum Cut} \label{sec:mincut-notemplatecut}

If no template crosses the cut, the challenge is to show how the minimum cut of the template graph relates to the minimum cut of the instantiation. Instead of the minimum cut value scaling with the parameters (as for the maximum all-$s$-$t$ flows), it is unaffected by the parameters.

The following statement implies that if no template crosses a cut, it can be found by looking at a single instance of some template (where all its child templates have been contracted). 

\begin{lemma}\label{lem:mincut-structure-nocross}
Consider a template graph that contains a minimum cut such that no template crosses the cut. Then, there is a minimum cut of $\mathcal{G}$ such that for every template $T$: either all vertices in instances of $T$ are on the same side of the minimum cut as the boundary vertices of the instance, or one side of the minimum cut is a subset of the vertices in exactly one instance of $~T$ (and contains no other vertices).
\end{lemma}
\begin{proof}
	The proof is by induction on the number of parameters. Consider a minimum cut $C$ of $\mathcal{G}$ where no template crosses the cut. 
	Consider a vertex $v$ that belongs to an instance $I$, where $v$ is on a different side of the cut $C$ as the boundary vertices of the instance $I$ (if no such vertex exists, all vertices are on the same side of the cut as the boundary vertices of the instances that contain them and we are done). There are two cases:
	
	\begin{description}
		\item [\textbf{Case 1.} \emph{The instance $I$ has at least one child $I'$.}] By assumption, this child $I'$ has its boundary vertices on the same side of the cut. Hence, all vertices in this child instance $I'$ must be on the same side of the cut as the boundary vertices of $I'$, since otherwise one side of the minimum cut would induce at least two connected components (which contradicts its minimality -- If one side of the cut induces at least two connected components, switching one such connected component to the other side reduces the cut value). Therefore, we can merge the child template (which instantiated $I'$) with its boundary vertices and obtain a parametric graph template with fewer parameters and the same minimum cut value. By induction, this graph has a cut with the desired properties. This cut gives us a cut in the original graph with the desired properties (by replacing each vertex with the set of vertices that were merged into it).
		\item [\textbf{Case 2.} \emph{The instance $I$ has no children.}] In this case, there cannot be any vertices on the same side of $v$ that are not in $I$. If there would be any such vertices, the cut partitions would induce more than two connected components, again contradicting optimality.
	\end{description}
	
\end{proof}

\Cref{lem:mincut-structure-nocross} leads to a recursive algorithm to compute a minimum cut for the case where no template crosses one of the minimum cuts. The construction uses a notion of induced parametric subgraphs:

The \emph{parametric subgraph $\mathcal{G}[T]$ of $\mathcal{G}=(G, \mathcal{T}, \mathcal{P})$ induced by a template $T$ and its boundary vertices} has as its template graph the subgraph of $G$ induced by the vertices in the template $T$ together with $T$'s boundary vertices. It contains all templates in $\mathcal{T}$ that are descendants of $T$ and a root template containing all vertices. All templates have the same corresponding parameters as they do in $\mathcal{P}$, except that the parameter of the root is $1$ (this means that the template $T$ is only repeated once in the parametric subgraph).
Merging the vertices in $ \mathcal{ G}[T]$ that correspond to boundary vertices of $T$ in $\mathcal{G}$ produces \emph{the parametric subgraph $ \mathcal{\hat G}[T]$ of $\mathcal{G}$ induced by the template $T$} and its \emph{merged} boundary vertices.

\paragraph{Algorithm}
Repeat the following until no template is left:
\begin{enumerate}
	\item Choose a template $T$ that has no child.
	\item Consider the parametric subgraph $ \mathcal{\hat G}[T]$ of $\mathcal{G}$ induced by the template $T$ and its \emph{merged} boundary vertices. Compute the minimum cut of its template graph, which equals its minimum cut (note that $ \mathcal{\hat G}[T]$ has a single template).
	\item After processing template $T$, merge all its vertices with its boundary vertices (in the main parametric graph template).
\end{enumerate}

\paragraph{Runtime} When we visit a template $T_i$, all its descendants have been replaced by a single edge or have been deleted. Therefore, the number of remaining vertices equals the number of vertices that belonged to $T_i$ in the original graph. There is at most one additional vertex that was not in $T_i$ in the first place, namely the one formed by merging the boundary vertices of the parent. The number of edges that remain is at most the number of edges that contain at least one vertex that belongs to $T_i$ plus the number of vertices that belong to $T_i$ (this accounts for the edges that might be added during the removal of the child templates).

Using Stoer-Wagner's algorithm~\cite{DBLP:conf/esa/StoerW94}, the minimum cut computations take $O(m \sum_i n_i +  n \sum_i n_i \log n)=O(mn + n^2 \log n)$ time. The cost to perform the minimum cut computations using the randomized Gawrychowski-Moses-Weiman algorithm~\cite{DBLP:conf/icalp/GawrychowskiMW20} is $O(\sum_i (m_i + n_i) \log ^ 2 n) = O(m \log ^ 2 n)$, where we used that every edge contributes to at most two terms in the sum and that the templates are connected.

\subsection{A Template Crosses the Minimum Cut}  \label{sec:mincut-templatecut}

If some template crosses the cut, we need to consider cuts where both sides of the cut contain only parts of a template. This might lead to complications if every instance of a template could behave completely differently. However, we can show that the cuts have a  symmetric structure, where (within some instance of a template) the templates that cross the minimum cut do so in a way where all instances of a vertex are on the same side of the cut. This follows by showing that the minimum cut equals a certain maximum $s$-$t$ flow in an induced parametric subgraph and applying the Edge Reweighting \Cref{thm:edge-reweighting}.

\begin{lemma}\label{lem:mincut-structure-cross}
	If there is a template that crosses some minimum cut $C$, then there is a minimum cut $C'$ of the following structure. There is a unique template $T$ such that one side of the cut is fully contained in a single instance $I$ of $~T$ and for every vertex $v$ that is contained in a \emph{child} of $~T$, all the instances of $v$ that are in $I$ are on the same side of the cut. 
\end{lemma}
\begin{proof}
Let $C$ be a minimum cut where some template crosses the cut $C$. Let $T'$ be a topmost template that crosses the cut $C$.  Let $I'$ be an instance of $T'$ with boundary vertices on both sides of the cut. The parent of $I'$ is either the root, or it has boundary vertices on the same side of the cut $C$ (otherwise $T'$ is not topmost). Let $I''$ be the instance found this way and $T''$ be its template. We continue to prove that the template $T''$ is the sought-after template $T$, by exhibiting a cut of the desired form (using \Cref{lem:stcut-structure}).

Let $A$ be the set of vertices on the side of the cut of the boundary vertices of $I''$ and $B$ be the set of vertices on the other side of the cut.
	Observe that the set $B$ is fully contained in the instance $I''$. This follows because $A$ and $B$ must each be a connected subgraph of the instantiation (and there is no way to enter the instance $I''$ without going through the boundary vertices).

Consider the parametric subgraph $\mathcal{G}[T']$ induced by the template $T'$ and its boundary vertices.
The sets $A$ and $B$ translate into sets of vertices $A'$ and $B'$ in the template graph $G'$ of the parametric subgraph $\mathcal{G}[T']$ (take for every vertex in the set the vertex that instantiated it). Any cut of $\mathcal{G}[T']$ that has one side of the cut containing $A'$ and the other side of the cut containing $B'$ can be turned into a cut of $\mathcal{G}$ of the same value.
 
In $\mathcal{G}[T']$, merge all the vertices in $A'$ into a single vertex $a$ and merge all the vertices in $B'$ into a single vertex $b$. Consider a minimum $a$-$b$ cut $C_{ab}$ of the merged parametric subgraph with the properties guaranteed by \Cref{lem:stcut-structure}. That is, all instances of the same vertex are on the same side of the cut $C_{ab}$. Undoing the merging gives the desired cut $C'$. This cut has the same value as $C_{ab}$ in $\mathcal{G}[T']$ and hence also $\mathcal{G}$ (Thus the cut $C_{ab}$ has value at least that of $C$). Additionally, since the cut $C$ corresponds to an $a$-$b$ cut in the merged graph, the minimum $a$-$b$ cut $C_{ab}$ has value at most $C$.

\end{proof}

\Cref{lem:mincut-structure-cross} together with \Cref{lem:stcut-reweighting} leads to a surprisingly simple algorithm to compute the minimum cut of a parametric graph template, in case some template crosses the minimum cut.
\paragraph{Algorithm}
For every template $T$:
\begin{enumerate}
	\item Apply Edge Reweighting to the parametric subgraph $ \mathcal{\hat G}[T]$ of $\mathcal{G}$ induced by the template $T$ and its \emph{merged} boundary vertices. 
	\item Compute the minimum cut of the reweighted graph.
\end{enumerate}
Return the smallest cut found overall.

\paragraph{Runtime} Every template is part of at most $h$ of the induced parametric subgraphs. Hence the runtime is $O(m h \log ^ 2 n)$ when using Gawrychowski-Moses-Weiman algorithm~\cite{DBLP:conf/icalp/GawrychowskiMW20} and $O(mn  h \log n + n^2 h \log n)$ when using Stoer and Wagner's algorithm~\cite{DBLP:conf/esa/StoerW94}.

\subsection{Characterizing the Parameter Space}

So far, we have focused on the viewpoint that the parameters are given and fixed, providing polynomial time algorithms, given a set of parameters. We can also characterize all cuts that can become minimum for certain functions of the parameters. Because Edge Reweighting transforms the parameters in an analytical way and the control flow of our algorithm is independent on the parameters, we have already shown how to effectively reduce the fully parametric case to the edge-weight parametric case for minimum cuts. 

The cuts that arise from the case where no template crosses the cut do not change with the parameters, and are the same for all parameters. There can be at most $O(n^2)$ such minimum cuts per template~\cite{DBLP:journals/jacm/KargerS96} and hence $O(n^3)$ such cuts overall. For the other cuts, we need to consider the shape of the edge reweighted graphs that arise.

When all parameters are restricted to have the same value $P$, then the edge reweighted problems that arise in the minimum cut computation are all of the following form: a weight $w(e)$ of edge $e$, whose deeper endpoint is at depth $i$ in the template tree, gets reweighted to $w'(e)=w(e) P^i$. These problems can be solved using Karger's parametric minimum cut approach~\cite{DBLP:conf/stoc/Karger16} in $O(n^3 m)$ time each, and they produce at most $O(n^3)$ cuts that could become minimum. Since we invoke this algorithm once for each template in the case where some template crosses the cut, the runtime is $O(n^4 m)$.

Another tractable case is when the template tree has height $h=1$ and there is a bounded number of templates $|\mathcal{T}|$. Then, the weight $w(e)$ of an edge that goes into template $T_i$ is reweighted to $w'(e)= w(e) P_i$. This can be seen as a linear combination of $|\mathcal{T}|$ terms (all but one have coefficient $0$). Hence, the runtime using Karger's approach~\cite{DBLP:conf/stoc/Karger16} is $O(mn^{1+|\mathcal{T}|})$ and there are $O(n^{1+|\mathcal{T}|})$ cuts that can become minimum.

\section{Template Subgraph Isomorphism} \label{sec:tree-si}

Having considered various cut and flow problems, we turn our attention to subgraph isomorphism problems. The goal is to detect if a certain \emph{pattern} graph can occur as a subgraph in some instantiation of a directed parametric graph template. The patterns we consider are fixed in advance and we focus on the case where the patterns are relatively small. Our goal is to derive an algorithm that is linear time if the number of vertices $k$ in the pattern is constant. 

First, we consider rooted tree-shaped patterns. Second, we show how finding rooted tree patterns in a parametric graph template can be used to find vertex-disjoint paths of bounded length. This can be thought of as a variant on a flow problem, where the paths corresponding to the flow are \emph{vertex disjoint} and have \emph{bounded length}.

Observe that solving the problem for the template graph does not solve the problem for the instantiation. In particular, a template graph that \emph{is not} a tree might instantiate a tree. Moreover, a template graph that \emph{is} a tree can instantiate a graph that is not a tree. 

Throughout this section, we consider \emph{template-acyclic} directed parametric graph templates. 
For now, we assume for simplicity that the parameters (except the root parameter) are all at least $k$ and that $h\leq k$. We remove these assumptions in \Cref{sec:si:generalp}.

\paragraph{Complexity} Even in graphs, tree subgraph isomorphism is hard in general. For undirected graphs, it is a generalization of the NP-hard \emph{longest path} problem~\cite{garey1979computers}. The hardness of deciding if a path of length $k$ exists follows by a reduction from the \emph{hamiltonian path problem}, for which there is a reduction from  the \emph{hamiltonian cycle} problem, one of Karp's NP-hard problems~\cite{DBLP:conf/coco/Karp72}. 

For directed acyclic graphs, there is a polynomial time reduction from a $3$-SAT variant (similar to the one used to show hardness of deciding if a certain number of disjoint paths of bounded length exist~\cite{DBLP:journals/networks/ItaiPS82}). Hence, the problem remains NP-hard in the directed acyclic case.
 
A naive recursive approach to decide if a pattern occurs in a graph is to recursively break up the pattern into parts. Then, record for every vertex $v$ all the subgraphs of the graph (which involve that vertex $v$) that are isomorphic to the parts of the pattern. Such a \emph{partial match} can then be used to build subgraphs that fit larger parts of the pattern. However, this is too slow in general, as there can be up to $n^k$ possible matches per vertex. Thus, we use Hierarchical Color Coding (see \Cref{sec:tech:color}) to reduce the state space of the possible partial matches. Recall that Hierarchical Color Coding assigns each level of the template tree (and in turn each template) a set of colors. 

\subsection{Tree Matching Algorithm}

The algorithm has two phases. The \emph{color coding phase} colors the vertices of the graph and guesses how the tree pattern matches onto the template tree as described in \Cref{sec:tech:color}. Then, for each of the colorings and mappings output by the first phase, the (deterministic) \emph{pattern matching phase} finds semi-colorful occurrences of the tree (i.e., colors are only allowed to repeat for instances of the same vertex). The algorithm recursively finds color sets of two subtrees and then checks if these can be safely combined. This check depends on the colors that appear in levels of the template tree that are at least as close to the root of the template tree as the root of the current pattern subtree. Colors that are deeper in the template tree are safe to combine because we can always create enough instances of the templates that contain these colored vertices to avoid collisions. 

In an instantiation of a parametric graph template, the \emph{spine} of a tree $\hat A$ rooted at $v$ is the set of vertices in $\hat A$ which belong to an instance of the template $T(v)$ or to a instance of a template that is an ancestor of $T(v)$. Intuitively, the spine contains vertices that occur at certain critical places in the template tree, where we cannot produce new instances of the vertices. Hence, it is important to avoid reusing colors that are in the spine. 
The algorithm produces for each vertex $v$ and for certain subtrees $\hat A$ of $A$ a list of sets of colors $C(\hat A, v)$ that the spine of the occurrences of $\hat A$, rooted at $v$, can have. 
 The pattern matching algorithm maintains the colors of the vertices in the spine of the current tree pattern and ensures that when combining two subtrees, they do not share any colors in the spine.

\paragraph{Algorithm}
As long as there is at least one edge in the current tree pattern $A$, proceed recursively:
\begin{enumerate} 
	\item Remove an edge $(u, v)$ incident to the root vertex $u$ of $A$. This creates two subtrees: a tree $A'$ rooted at $u$ (i.e., the tree $A$ without the subtree rooted at $v$) and a tree $A''$ rooted at $v$ (i.e., the subtree of $A$ that is rooted at $v$).
	\item Recursively invoke the pattern matching algorithm for the two resulting subtrees $A'$ and $A''$.
	\item For each edge $(x, y)$ in the template graph $G$, combine the lists of color sets $C(A',x)$ and $C(A'', y)$ that we obtained recursively. Consider a color set $c'$ in $C(A',x)$ and a color set $c''$ in $C(A'', y)$. Let $c_{\text{low}}$ be the set of colors that correspond to templates that are in a deeper level of the template tree than the template $T(x)$. 
	\begin{description}
	\item [Case 1: \emph{ $T(x)$ is a parent of $~T(y)$.} ]If $c'\setminus c_{\text{low}}$ and $c''\setminus c_{\text{low}}$ are disjoint, add $(c' \cup c'') \setminus c_{\text{low}}$ to the list of colors $C(A, x)$. 
	\item [Case 2: \emph{$T(x)$ is not a parent of $~T(y)$}.] If $c'$ and $c''$ are disjoint, add $c' \cup c''$ to the list $C(A, x)$. 
	\end{description}
\end{enumerate}
Once the tree $A$ has a single vertex $v$ (and no edge left), we reach the \emph{base case}: for each vertex $x$ in the graph $G$, add the set containing just the color of $x$ to the list $C(A, x)$.

\begin{figure}[b]
\includegraphics[width=0.8\linewidth]{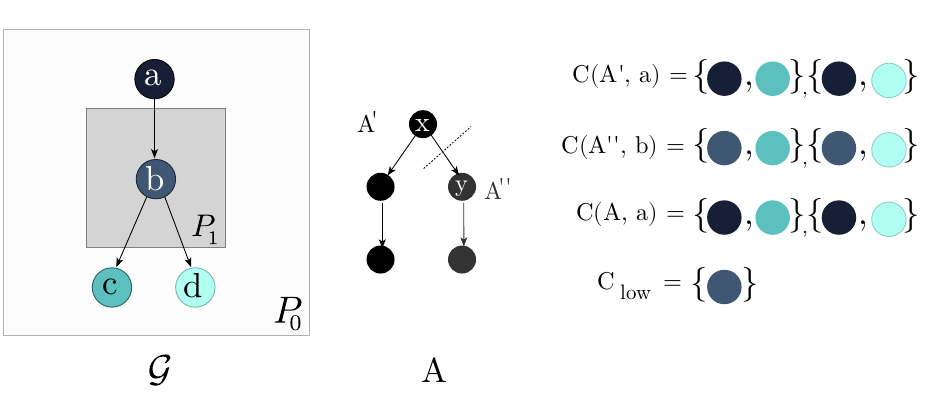}
\caption{Illustration of an intermediate step of the tree matching algorithm for the tree pattern $A$ and the hierarchically color coded $\mathcal{G}$. The figure shows Case 1 in step (3). The spine of any instance of $A'$ rooted at $a$ does not contain any instance of $b$, because $b$ is in a deeper level of the template tree as $a$.}\label{fig:treesi}
\end{figure}

See \Cref{fig:treesi} for an illustration of a step of the algorithm.

\paragraph{Runtime} 

	The first step is to bound the size of the lists $C(A, x)$. We observe that if a tree $A$ has $i$ vertices, the lists $C(A, x)$ are subsets of $k$ colors with at most $i$ elements. There are $\sum_{j=0}^{j=i} {k \choose j}$ such sets.
	Let $t(i)$ be the runtime given that the current tree has $i$ vertices.
	For some constants $c_1$ and $c_2$, the runtime $t(i)$ follows the recurrence:
	\begin{align*}
	t(i) &\leq 
	\begin{cases}
		c_1 m & \text{if $i=1$} \\
		 c_2 \ m \ i \left (\sum_{j=0}^{j=i} {k \choose j} \right )  + t(i-r) + t(r) & \text{else, \enspace for some $r$ with $0<r<i$ \enspace .}
	\end{cases}
	\end{align*}
	To bound the runtime, think of the recurrence as a binary recursion tree, where a node corresponds to an invocation of the procedure (each nodes stores the number of vertices of the tree it is invoked on) and each node is connected to the two recursive calls it makes. Observe that the recursion tree has $k$ leaves (because the sum of the arguments of the two recursive call always sums to the argument of the current call). Moreover, for every $i$, the term $t(i)$ appears at most once along every root to leaf path in the recursion tree, because the arguments are strictly decreasing along such a path. Therefore, for every $i$, the term $i \sum_{j=0}^{j=i} {k \choose j}$ appears at most $k$ times during the recursion. Hence, the runtime is bounded by $$O\left(m k^2 \sum_{i=0}^{k} \left (\sum_{j=0}^{j=i} {k \choose j} \right) \right) = O\left(m k^2 \sum_{i=0}^{k} \left ( (k-i) {k \choose i} \right) \right)  = O(m k^3 2^{k}) \enspace ,$$ where the first equality follows by inspection of the double-sum and the second equality follows from the binomial theorem. 
	
To remove the assumption that $h\leq k$, we can modify Hierarchical Color Coding by treating levels of the template tree \emph{modulo $k+1$} (i.e., templates at depth $i$ and $i+k$ have the same set of colors). This suffices, because all we need to know for the algorithm is whether a given color originates from deeper levels in the template tree or not (and any occurrence of the tree $A$ is contained in at most $k$ levels of the template tree).

Together with the bounds on Hierarchical Color Coding (\Cref{obs:colorcode1}, \Cref{obs:colorcode2}), this implies a runtime of $O(mk^4 2^k e^k 3^k \log n)$ to decide tree subgraph isomorphism with high probability when all parameters $P_i$ satisfy $P_i\geq k$.

\subsection{Handling General Parameters}\label{sec:si:generalp}

So far, we assumed that the parameters $P_i$ all satisfy $P_i\geq k$. This allowed us to disregard the number of times we reuse a color. We can remove this assumption with some additional bookkeeping. Naively, one could count for every template how many times it is used. But this would lead to an $\Omega(n^k)$ increase in runtime. Instead, we observe that when the parameters are larger than $k$, we do not need to know exactly how big they are. This means that we can group the templates into at most $k$ groups based on their parameters: a group for each parameter less than $k$ and one group for parameters at least $k$. Then, similarly to how we restricted the color-coding to be hierarchical, we now make it \emph{parameter aware}:

Guess to which group a vertex in the pattern belongs and use as many colors for a group as there are vertices of the pattern in that group. There are at most $k^k$ ways of assigning the vertices of the pattern to the groups.
Keeping track of how many times a color has been used suffices to ensure that we respect the parameters. This is because we search for semi-colorful occurrences, so all vertices of the same color must be instances of the same vertex and therefore belong to an instance of the same template. The bookkeeping takes  $O(k^k)$ additional state. 

\subsection{Proof of Correctness}

Before we turn to proving the correctness of the algorithm, we show a property about paths in template-acyclic graphs. We will use the following \Cref{cor:acyc-paths} to show that the algorithm finds all semi-colorful occurrences.

\begin{lemma}
	[Template-Acyclic Paths]\label{cor:acyc-paths}
	In an instantiation of a template-acyclic parametric graph template,
	consider a path $P_{uv}$ from vertex $u$ to vertex $v$
	and a path $P_{wz}$ from vertex $w$ to vertex $z$ where $T(v)=T(z)$. If $~T(v)=T(u)$, or $T(v)$ is an ancestor of $~T(u)$, then if $u$ and $w$ belong to the same instance of the same template, then $v$ and $z$ do as well.
\end{lemma}
\begin{proof}
The proof is by strong induction on the sum of the length of the two paths. If the lengths are $0$, then $u=v$ and $w=z$ and the statement follows immediately. Otherwise, assume as the induction hypothesis that the statement holds for all pairs of paths whose sum of the path lengths are shorter that the sum of the length of $P_{uv}$ and $P_{wz}$.

In a template-acyclic graph, it is impossible (by definition) that there is a path from an instance of a template to another instance of the same template. This proves the case where $T(u)=T(v)$. For the case where $T(v)$ is an ancestor of $T(u)$, recall that paths in an instantiation of a parametric graph template correspond to walks in the template tree. Now, consider the first edge $(x, y)$ in the path $P_{uv}$ where $y$ belongs to an instance of $T(v)$ and the first edge $(x', y')$ in the path $P_{wz}$ where $y'$ belongs to an instance of $T(v)$. By construction, $T(x)=T(x')$ where this template is a child of $T(v)$. Therefore, by induction, $x$ and $x'$ are in the same instance of the same template. Now, since edges can only cross one template instance boundary at a time, $y$ and $y'$ must also be in the same instance of the same template. Apply the induction hypothesis for the path $P_{yv}$ from $y$ to $v$ and the path $P_{y'z}$ from $y'$ to $z$ to conclude the proof.
\end{proof}

\begin{corollary}[Converging Paths]\label{cor:conv-paths}
In an instantiation of a template-acyclic parametric graph template,
	consider an instance $u'$ of vertex $u$ and two instances $v'$ and $v''$ of the same vertex $v$ where either $T(u)=T(v)$ or $T(v)$ is an ancestor of $~T(u)$. If there is a path from $u'$ to $v'$ and a path from $u'$ to $v''$, then $v'=v''$.
\end{corollary}
\begin{proof}
Apply the template-acyclic paths \Cref{cor:acyc-paths} for the path $P_{u'v'}$ and the path $P_{u'v''}$
\end{proof}

\begin{lemma}
	[Completeness]
	If there is a semi-colorful occurrence of the tree $A$ rooted at $x$ using the colors in the set $c$ on its spine, then the set of colors $c$ is in the list $C(A, x)$.
\end{lemma}
\begin{proof}
The proof is by structural induction along the recursive structure of the algorithm.
The base case of the algorithm is clearly correct.
Consider a semi-colorful occurrence of $A$ rooted at $x$ using the colors in the set $c$. Split the tree $A$ into $A'$ and $A''$ as in the algorithm. This yields a semi-colorful occurrence of tree $A'$ rooted at $x$ using a set of colors $c'$ on its spine and a semi-colorful occurrence of $A''$ rooted at a vertex $y$ using a set of colors $c''$ on its spine, where there is an edge $(x, y)$. The induction hypothesis is that the algorithm finds that $c'$ is in $C(A', x)$ and that $c''$ is in $C(A'', y)$. 

\begin{description}
	\item [Case 1. $T(x)$ is a parent of $T(y)$. ] 
	
We need to show that $(c' \cup c'') \setminus c_{\text{low}}$ is the set of colors on the spine of the occurrence of $A$ and that $(c'\setminus c_{\text{low}})$ and  $(c''\setminus c_{\text{low}})$ are disjoint.

In this case, the spine of the occurrence of $A$ contains the spine of the occurrence of $A'$ and the spine of the occurrence of $A''$, without the vertices that belong to $T(y)$. By the induction hypothesis, the sets $c'$ and $c''$ contain colors of the vertices in the spines of the occurrence of $A'$ and $A''$, respectively. Moreover, by definition the color set $c_{\text{low}}$ contains the colors that occur in the template $T(y)$ and no colors that occur in the spine of the occurrence of $A$. Hence, $(c' \cup c'')\setminus c_{\text{low}}$ is indeed the set of colors that occur on the spine of the occurrence of $A$.

To prove that $(c'\setminus c_{\text{low}})$ and  $(c''\setminus c_{\text{low}})$ are disjoint, assume otherwise (for contradiction). Since the occurrence of $A$ is semi-colorful, there must be a vertex $v'$ in the occurrence of $A'$ and a vertex $v''$ in the occurrence of $A''$ that are instances of the same vertex $v$. Since $c_{\text{low}}$ contains the colors in $T(y)$ (and using the induction hypothesis), $(c'\setminus c_{\text{low}})$ and  $(c''\setminus c_{\text{low}})$ must intersect in a color that is in $T(x)$ or an ancestor of $T(x)$. Moreover, there is a path from $x$ to $v'$ (along the occurrence of $A'$) and a path from $x$ to $v''$ (using the edge $(x, y)$ and then along the occurrence of $A''$). Hence, by Converging Paths (\Cref{cor:conv-paths}), $v'=v''$, which contradicts that $A$ is a tree. 
	\item [Case 2. $T(x)=T(y)$ or $T(x)$ is a child of $T(y)$.]
	We need to show that $c' \cup c''$ is the set of colors in the spine of the occurrence of $A$ and that $c'$ and $c''$ are disjoint.

Because $T(x)=T(y)$ or $T(x)$ is a child of $T(y)$, the spine of the occurrence of $A$ contains the spine of the occurrence of $A'$ and the spine of the occurrence of $A''$. Therefore, the set $c' \cup c''$ is indeed the colors of the spine of the occurrence of $A$.

To show that $c'$ and $c''$ are disjoint, assume otherwise (for contradiction). Then, because the occurrence of $A$ is semi-colorful, there is a vertex $v'$ in the spine of the occurrence of $A'$ and a vertex $v''$ in the spine of the occurrence of $A''$ that are both instances of the same vertex $v$. Because $T(x)=T(y)$ or $T(x)$ is a child of $T(y)$, we conclude that $v'$ and $v''$ must either belong to $T(x)$ or an ancestor of $T(x)$. Moreover, there is a path from $x$ to $v'$ and $x$ to $v''$ (along the occurrence of the tree $A'$ and the edge $(x, y)$ followed by a path in the occurrence of $A''$, respectively). Hence, by Converging Paths (\Cref{cor:conv-paths}), $v'$ and $v''$ must be the same vertex, which again contradicts that $A$ is a tree.
\end{description}
\end{proof}

\begin{lemma}[Safety]
	If the list $C(A, x)$ contains the color set $c$, then there is an occurrence of the tree $A$ rooted at $x$ using exactly the colors from the set $c$ on its spine.
\end{lemma}
\begin{proof}[Proof]

The idea is again to proceed by induction. We get two subtrees $A'$ and $A''$ as before where $A'$ is rooted at $x$ and $A''$ at some child of $x$. The induction hypothesis guarantees us that there are occurrences of $A'$ and $A''$ using colors $c'$ and $c''$ on their spines, respectively. We need to prove that there is an occurrence of $A$ using the colors $c$ on its spine as constructed in the algorithm. This involves showing that either the vertices in the two subtrees are distinct, or mapping the occurrences onto other vertices so that this is the case.

Vertices in the spine of the occurrence of $A$ cannot be confused (their occurrences must be distinct) as their colors are distinct.

Outside of the spine of the occurrence of $A$, we can always combine instantiations of $A'$ and $A''$ by choosing different instantiations. Specifically, if there is a potential clash between instances of a vertex $v$ in the occurrences of $A'$ and $A''$, consider the least common ancestor $T'$ of $T(v)$ and $T(x)$. Let $T''$ be the child of $T'$ that contains $v$. There is such a template because $v$ is not in the spine of $A$ and $T(v)$ is the descendant of some template in the spine. 
Choose two different instances for the part of $A'$ that is in $T''$ and the part of $A''$ that is in $T''$. This works because $T''$ is not in the spine of $A$, and since there are always enough instances (because we count the number of times a color is reused). 
\end{proof}

This concludes our proof of \Cref{thm:isomorphs}. We now explore a variation of the algorithm for a related problem.

\subsection{Vertex-Disjoint Bounded Length Paths}\label{sec:tree-si-vdp}

We want to decide if $k$ vertex-disjoint $s$-$t$ paths of length exactly $L$ occur in some instance of the parametric graph template (note that the paths can share the vertices $s$ and $t$).

\paragraph{Complexity}
For integer weights, the maximum flow problem corresponds to finding a maximal set of edge-disjoint $s$-$t$ paths. These paths can have unbounded length. When we restrict the length of these paths, the problem becomes NP-hard even when the length bound is $O(1)$, and similarly if the paths are required to be vertex-disjoint~\cite{DBLP:journals/networks/ItaiPS82}. 

\paragraph{Algorithm}
Run the tree subgraph isomorphism algorithm for a tree $T$, which has $k$ leaves and where every leaf is connected to the root with a length $L-1$ path. Moreover, modify the pattern matching phase as follows:

Whenever a leaf $v$ of $A$ is reached, for each vertex $x$ in the graph, only add the color of $x$ to $C(v, x)$ if there is an edge from $x$ to the ``sink'' vertex $t$. To decide if there are $k$ vertex-disjoint paths of length exactly $L$, check if $C(A, s)$ contains at least one element. If any of the colorings from the color coding phase return an occurrence, there is an occurrence.

\paragraph{Runtime}
The runtime equals to that of the tree subgraph isomorphism algorithm for $kL$ vertices. It is polynomial in $n$ and $m$ as long as $kL=O(1)$. In particular, we obtain the following bound: 

\begin{corollary}\label{thm:tree-si-bound}
Deciding if there are parameters such that there are $k$ vertex-disjoint paths of length exactly $L$ in a template-acyclic parametric graph template rooted at vertex $v$, for all $v$, takes  $(k L)^{O(kL)} m \log n$ time. The algorithm is correct with high probability.
\end{corollary}

The algorithm can also be adapted further to detect occurrences of $k$ vertex-disjoint $s$-$t$ paths of length \emph{at most} $L$:

\paragraph{Algorithm}
First, count the number $k'$ of edges between $s$ and $t$. Then, delete these edges from the template graph and run the algorithm to find $k-k'$ disjoint paths of length $L$, with the following modification. We introduce the possibility of an ``early base case'' in the pattern matching phase: say the current tree $A$ consists of a path of any length and edge $(x, y)$ is being considered where $y=t$; then, add the color of $x$ to $C(A, x)$. 

This works because the modified pattern matching phase finds vertex-disjoint paths of length at least $2$ and at most $L$. We separately count the paths of length $1$.

\begin{corollary}
\label{thm:tree-si-bound}
Deciding if there are parameters such that there are $k$ vertex-disjoint paths of length \textbf{at most} $L$ in a template-acyclic parametric graph template rooted at vertex $v$, for all $v$, takes  $(k L)^{O(kL)} m \log n$ time. The algorithm is correct with high probability.
\end{corollary}

\section{Template Discovery}\label{sec:autotemplate}

We turn to studying the fundamental question of ``Can we discover a concise parametric graph template that instantiates a given graph?''. Besides being an interesting question about the tractability of our model, it might be necessary to answer this question in settings where the parametric graph template representation is not known a priori (like biological settings).

We show that given a graph $G'$, we can enumerate all parametric template graphs that instantiate the graph $G'$. Our algorithms are quasi-polynomial time when the template graphs are required to have a constant number of vertices that are boundary vertices of the children. The runtime is improved on graphs of bounded diameter or low treewidth, a well-studied structural graph parameter. The algorithm exploits several recent advances in graph isomorphism algorithms.

We denote the number of vertices and edges of the target graph by $N$ and $M$, respectively.
Throughout, we assume w.l.o.g. that the subgraphs of the template graph induced by any one template is connected. If the template graph is disconnected, then the templates can be split until this condition holds.

\paragraph{Treewidth and Tree Decomposition}

Intuitively, treewidth measures the ability of a graph to be separated recursively into subgraphs only overlapping in a small number of vertices. Many important graphs have a small treewidth, for example planar graphs~\cite{DBLP:conf/soda/Eppstein95} (or bounded genus graphs~\cite{DBLP:journals/algorithmica/Eppstein00}) of bounded diameter, or graphs arising from certain electric circuits~\cite{DUFFIN1965303}. Although deciding if a graph has a certain treewidth is NP-hard~\cite{doi:10.1137/0608024}, there are efficient algorithms for deciding if a graph has \emph{bounded treewidth}~\cite{DBLP:journals/jal/BodlaenderK96, DBLP:conf/stoc/Bodlaender93, DBLP:conf/focs/BodlaenderDDFLP13}.
Many NP-hard problems, such as subgraph isomorphism~\cite{DBLP:conf/soda/Eppstein95}, chromatic number~\cite{DBLP:journals/dam/ArnborgP89}, and graph isomorphism~\cite{DBLP:conf/focs/LokshtanovPPS14}, become polynomial time solvable on bounded-treewidth graphs. Other problems can be efficiently approximated on such graphs~\cite{DBLP:journals/jacm/Baker94}.

A tree decomposition of a graph $G'$ consists of a binary \emph{decomposition tree} $\mathcal{A}$ whose nodes are sets of vertices of $G'$ (to avoid confusion, we use the term \emph{nodes} when referring to the vertices of the decomposition tree). These nodes satisfy the following two conditions:
\begin{enumerate}
	\item Each vertex of $G'$ occurs in a connected (and nonempty) subtree of the decomposition tree $\mathcal{A}$.
	\item For each edge $(u, v)$ of $G'$, there is a node of $\mathcal{A}$ that contains both endpoints $u$ and $v$.
\end{enumerate}
The width of the decomposition tree is one less than the size of the largest node. This correction by one is to ensure that trees have treewidth $1$. The \emph{treewidth} $\tau$ is the smallest width of any decomposition tree. 
Finding a binary decomposition tree with $O(N)$ nodes and width $\tau$ takes $O(g(\tau)N)$ time for some function $g$~\cite{DBLP:journals/jal/BodlaenderK96, DBLP:conf/stoc/Bodlaender93}.
%

\paragraph{Graph Isomorphism}

The last five years have seen major breakthroughs in the graph isomorphism problem. The fastest general-purpose algorithm runs in quasi-polynomial time $N^{\text{polylog} (N)}$~\cite{DBLP:conf/stoc/Babai16}. For graphs of maximum degree $d$, this can be improved to $N^{\text{polylog} (d)}$ time~\cite{DBLP:conf/focs/GroheNS18}. Interestingly, the problem can be solved in  polynomial time for bounded treewidth, namely $2^{O(\tau^5 \log \tau)}N^5$ time~\cite{DBLP:conf/focs/LokshtanovPPS14}.

\subsection{A Template Discovery Algorithm}

The algorithm recursively tries to construct the templates in a top-down manner (along the template tree). It uses graph isomorphism queries to discover subgraphs that belong to the same template.

\paragraph{Algorithm}
Enumerate all sets of potential boundary vertices of the children $B$. For each such set $B$, remove the set $B$ from the graph and continue as follows:
\begin{enumerate}
	\item Group the resulting connected components such that each group consists of isomorphic graphs. 
	\item Create a root template by including the boundary vertices of the children $B$ and all groups that consist of a single graph.
	\item For each group of at least two isomorphic connected components, recursively create a template graph and assign it as a child template, connecting it to the vertices in $B$ appropriately.
\end{enumerate}
To solve the graph isomorphism questions, we can use the graph isomorphism for bounded treewidth graphs and the aforementioned quasi-polynomial time algorithm for general graphs~\cite{DBLP:conf/stoc/Babai16}. 

\begin{proof}[Proof Of \Cref{thm:autotemplate}]
Because every child's parameter is at least $2$, each recursive subproblem is on a graph that has at most half the number of vertices. Moreover, this spawns at most $N^{\beta+1}$ subproblems. Let GI($N$) denote the runtime to decide graph isomorphism and let $t(N)$ denote the runtime of the template discovery algorithm. For some constants, $c_1$, $c_2$, we get the recurrence:
\begin{align*}
	t(N) \leq 
\begin{cases}
c_1 & \hfill \text{if $n \leq 1$ ,}\\
N^{\beta+1} t(\frac{N}{2}) + c_2 N \cdot \text{GI}(N) & \hfill \text{otherwise \ .}
\end{cases}
\end{align*}
The runtime is dominated by the time spent towards the leaves of the recursion and is bounded by $2^{O(\tau^5 \log \tau)}N^{O(\beta \log N)}$ for target graphs of bounded treewidth and $N^{ \beta \ \text{polylog}( N )}$ for general target graphs.
\end{proof}

Note that the runtime does not depend on the size of the template, but only on $\beta$. It is especially well-suited when the smallest parameter $P_i$ is polynomial in $N$, in which case the depth of the recursion is constant and the runtime becomes $2^{O(\tau^5 \log \tau)}N^{O(\beta )}$.

\section{Discussion and Open Problems}

In this work, we explore the notion of structural parameterization in graphs. Specifically, we show that hierarchically repeating graph templates can be recovered from existing graphs and algorithmically studied. Moreover, our model leads to polynomial time algorithms for many important graph problems, notably maximum $s$-$t$ flows, minimum cuts, and subgraph isomorphism for bounded-size tree pattern graphs. 
This indicates a promising direction in scaling algorithms to graph classes that exhibit this hierarchically self-repeating behavior.

The parametric graph template model poses further questions: which other algorithms can be expressed as a function of the underlying template graphs? Can the model be generalized further to support wider forms of parametric structures (for example, allowing edges to connect instances of different templates)? It would also be interesting to improve the presented bounds, such that the time complexity on parametric graph templates matches the one of solving the problem on the template graph. 
Lastly, one could investigate if it is possible to derive a template discovery algorithm that is fixed-parameter tractable (for example in the treewidth and the maximum boundary nodes of any child templates $\beta$).

\bibliographystyle{alpha}
\bibliography{dapp-bib,relatedwork,isomorph-apps,isomorphs, graph-grammar}

\end{document}